\documentclass[a4paper,fleqn]{cas-dc}

\usepackage[authoryear,longnamesfirst]{natbib}

\usepackage{amsmath,amssymb,amsfonts}
\usepackage{algorithmic}
\usepackage{graphicx}
\usepackage{booktabs}
\usepackage{tabularx}
\usepackage{siunitx}
\usepackage{textcomp}
\usepackage{xcolor}
\usepackage{hyperref}
\usepackage{xurl}
\usepackage{tcolorbox}
\usepackage{pifont}
\usepackage{balance}

\def\tsc#1{\csdef{#1}{\textsc{\lowercase{#1}}\xspace}}
\tsc{WGM}
\tsc{QE}

\def\BibTeX{{\rm B\kern-.05em{\sc i\kern-.025em b}\kern-.08em
    T\kern-.1667em\lower.7ex\hbox{E}\kern-.125emX}}

\newcommand{\rqn}[1]{RQ\texorpdfstring{\textsubscript{#1}}{#1}}
\newcommand{\code}[1]{\texttt{#1}}

\tcbset{arc=0mm,boxrule=0.1mm,before skip=1mm,after skip=1mm,top=0.5mm,bottom=0.5mm,left=1mm,right=1mm,colback=white,colframe=black,fontupper=\footnotesize}

\begin{document}
\let\WriteBookmarks\relax
\def\floatpagepagefraction{1}
\def\textpagefraction{.001}

\shorttitle{Mining for Awareness Cost in IaC Artifacts}    

\shortauthors{Feitosa et al.}

\title [mode=title]{Mining for Cost Awareness in the Infrastructure as Code Artifacts of Cloud-based Applications: an Exploratory Study}

\author[1]{Daniel Feitosa}[orcid=0000-0001-9371-232X]
\cormark[1]
\ead{d.feitosa@rug.nl}
\ead[url]{https://feitosa-daniel.github.io}
\credit{Conceptualization, Methodology, Software, Validation, Formal analysis, Data Curation, Writing - Original Draft, Supervision}

\author[1]{Matei-Tudor Penca}
\ead{matei.penca1@gmail.com}
\credit{Software, Formal analysis, Data Curation}

\author[1]{Massimiliano Berardi}
\ead{massimiliano.berardi93@gmail.com}
\credit{Software, Formal analysis, Data Curation}

\author[1]{Rares-Dorian Boza}
\ead{raresboza@gmail.com}
\credit{Software, Formal analysis, Data Curation}

\author[1]{Vasilios Andrikopoulos}[orcid=0000-0001-7937-0247]
\ead{v.andrikopoulos@rug.nl}
\ead[url]{https://vandriko.github.io}
\credit{Conceptualization, Methodology, Validation, Writing - Original Draft, Supervision}

\affiliation[1]{organization={Bernoulli Institute for Mathematics, Computer Science and Artificial Intelligence, University of Groningen}, country={The Netherlands}}

\cortext[1]{Corresponding author}

\begin{abstract}\mbox{\hspace{-3pt}}
\textbf{Context:} Cloud computing's rise as the primary platform for software development and delivery is largely driven by the potential cost savings.
However, it is surprising that no empirical evidence has been collected to determine whether cost awareness permeates the development process and how it manifests in practice.
\newline
\textbf{Objective:} This study aims to provide empirical evidence of cost awareness by mining open source repositories of cloud-based applications.
The focus is on Infrastructure-as-Code artifacts that automate software (re)deployment on the cloud.
\newline
\textbf{Methods:} A systematic examination of $152\,735$ repositories yielded $2\,010$ relevant hits.
We then analyzed 538 relevant commits and 208 relevant issues using inductive and deductive coding and corroborated findings with discussions from Stack Overflow.
\newline
\textbf{Results:} The findings indicate that developers are not only concerned with the cost of their application deployments but also take actions to reduce these costs beyond selecting cheaper cloud services.
We also identify research areas for future consideration.
\newline
\textbf{Conclusion:} Although we focus on a particular Infrastructure-as-Code technology (Terraform), the findings can be applicable to cloud-based application development in general. 
The provided empirical grounding can serve developers seeking to reduce costs through service selection, resource allocation, deployment optimization, and other techniques.
\end{abstract}


\maketitle

\section{Introduction}
\label{sec:intro}

Cost reduction is one of the main drivers of cloud adoption \citep{Andrikopoulos2013}. 
Cost savings for the cloud consumers accrue due to two phenomena. 
First, access to any kind of computational resources (both hardware and software) is on-demand and is being billed utilities-style \citep{mell2011nist}. 
This means that scaling up and down the amount of these resources to meet the needs of the current load leads to higher efficiency in comparison with a fixed infrastructure such as e.g.\ in a traditional data center. 
Compounding this, there is also no need for upfront capital expenses for the acquisition of these resources, e.g. to cope with unforeseen demand. 
This means a nearly complete transfer of the focus from the management of capital expenses to operational ones, also known as the CAPEX-to-OPEX shift \citep{armbrust2010view}. 
Second, and on top of that, the economies of scale realized by the cloud service providers, and especially by the ones such as Amazon Web Services, Microsoft Azure, and Google Cloud Platform, collectively known as \emph{hyperscalers} due to their ability to enable scaling to virtually infinite levels of demand, allow these providers to offer access to these resources for almost (always) declining prices \citep{harms2010economics}.
This makes cloud computing very attractive to all kinds and sizes of organizations and enterprises.

Further testimony to the importance of cost for adopters of cloud computing is the amount of related work on areas investigating how to minimize and/or manage this cost.
This can be achieved, for example, from the perspective of cloud consumers through optimal cloud service provider selection \citep{tricomi2020optimal,hosseinzadeh2020service}, and from the perspective of the providers by means of optimized task scheduling \citep{arunarani2019task} or other profit optimization techniques such as energy consumption minimization \citep{cong2020survey}.
Whats more, every cloud service provider offers in one form or another cost calculator tools, allowing their users to get a quote on their service consumption based on their foreseen computational, storage, network etc.\ needs.
Consulting on cloud cost management has developed into its own line of business, with even cloud service providers themselves offering such services to their users.
In all cases, cost is referring to the monetary expenses of hosting and running software in one of the cloud deployment models as defined by NIST \citep{mell2011nist}.

However, and to the extent of our knowledge, no empirical evidence exists on whether and in what form the cost of cloud-based software projects is discussed among the involved developers.
While on the surface such concerns appear to be outside of the remit of software development per se, the situation in practice is quite different.
The fact, for example, that the DevOps paradigm became popular and widely adopted almost in the same timeframe as cloud computing hints that software developers cannot easily ignore the operational aspects of the code they produce, including its cost.
Furthermore, the aforementioned utilities-like billing of cloud services means that developers are now in a position to be held effectively accountable for the generated revenue of the software that they produced, deployed, and ran on the cloud infrastructure.
As such, and further bolstered by the amount of anecdotal evidence, we do expect software developers to be actively aware and concerned about the cost of their software, and we set out to collect evidence of this.

The objective of this study is therefore clear: \emph{to examine to what extent software developers are aware of the cost of deploying and operating cloud-based software, and what kind of concerns and action initiatives they are having about it.}
We choose MSR (Mining Software Repositories) as the means to answer this question empirically.
Among its other uses, MSR allows to empirically study otherwise subjective or external phenomena in combination with (or through) large-scale systematic mining of development artifacts.
Fields such as green software engineering \citep{hindle13greemining,pereira21greenhub}, risk assessment \citep{choetkiertikul15issuerisks,costa2017rapid,choetkiertikul18delivery}, and software classification \citep{howard13similarwords,leclair18neuraltext,sas22antipatterns} have advanced noticeably due to MSR.
In a similar fashion to these works, we hypothesize that \emph{the amount and diversity of costs-related information in cloud-based software project repositories is sufficient to produce meaningful insights}.

Cloud-based application development, however, covers a very wide range of application types and development activities, and this creates a question of scope in this study.
Selecting for a specific programming language or ecosystem as in other MSR studies does not produce meaningful results here since these are orthogonal concerns to the use of cloud infrastructures.
Instead, we scope our search for evidence to \emph{cloud orchestrator artifacts} included in open source projects.
Cloud orchestrators are \emph{Infrastructure as Code} (IaC) solutions that provide an abstraction layer over the self-service management APIs of the various cloud service providers, with the intention of flattening out the differences between them \citep{deCarvalho2020}.
This is usually achieved by means of \emph{descriptor files}, i.e., configuration files that when interpreted by the orchestrator ensure that both the underlying infrastructure is made available, and the tasks required for the (re)deployment of software on this infrastructure is executed correctly.
Descriptor files are usually semi-structured documents in a machine-readable format that is easy to process such as YAML or JSON, and like any other configuration files they are added to code repositories to be managed by the respective version control system. 
Orchestrators are either (cloud service) provider-specific, such as Amazon Web Services CloudFormation, or provider-agnostic, such as Terraform, Cloudify, Apache Heat, and others as discussed for example in \cite{tomarchio2020cloud}.
For reasons that will be discussed in Section~\ref{sec:design}, we focus our work specifically on Terraform artifacts.

In summary, this paper aims to report on the first work that attempts to perform cost awareness mining for cloud-based application development, starting with IaC artifacts.
In addition, we fortify the findings of the mining process by also investigating to what extent the identified related topics are discussed on a popular developer forum, Stack Overflow\footnote{\url{https://stackoverflow.com/}}, in relation to the same type of artifacts.
The analysis of the involved posts confirms the fitness of our result to purpose.
The contributions of this work can therefore be summarized as follows:
\begin{itemize}
    \item we collect and present \emph{empirical evidence} of the existence of cost-related information pertinent to Terraform artifacts as it appears in (open) source code repositories;
    \item we triangulate and augment this information extraction by identifying Stack Overflow posts where pertinent discussions take place;
    \item we make publicly available a \emph{curated dataset} of the artifacts identified through this evidence-collection process and the scripts we used during the data collection, see~\cite{supplementary_material};
    \item we define a set of \emph{actionable items for future research on cost awareness} based on our preliminary analysis of this dataset.
\end{itemize}

The rest of this paper is structured as follows.
Some related works are presented in Section~\ref{sec:background}.
Section~\ref{sec:design} discusses the study design, including the definition of research questions to investigate.
Section~\ref{sec:results} presents our findings, and Section~\ref{sec:triangulation} discusses our effort to triangulate them through developers' interactions on a public forum.
Section~\ref{sec:implications} offers a discussion on the implications of these findings for practitioners and researchers, including the formulation of a research agenda for future work.
Finally, Sections~\ref{sec:ttv} and~\ref{sec:conclusions} close this paper with a reflection on the threats to validity to this study, and a summary of its main findings, respectively.

\section{Related Work}
\label{sec:background}

As mentioned in the previous section, and to the best of our knowledge, there is no existing study gathering empirical evidence about how developers deal with the cost of deploying cloud-based or otherwise software, and definitely none collected through repository mining.
The closest works in spirit in this direction are instead studies on mining energy consumption awareness on the developer's side such as the work by \cite{Moura2015} and \cite{Bao2016}. 
Other works such as the one by \cite{pinto2014mining} on the same topic, or \cite{das2016quantitative} analyzing documented performance-related issues can be also considered somewhat related to ours.

However, that is not to say that there are no research efforts for supporting the management of cost in such systems.
In fact, that is a flourishing line of research approached from different perspectives.
Despite being recognized early on as a major concern when migrating existing systems to the cloud \citep{Andrikopoulos2013}, for example, and being a crucial component in many migration support approaches \citep{jamshidi2013cloud}, estimating the cost of deploying and running software in the cloud remains an open research challenge \citep{shuaib2019adopting}.
Cost of deployment and operation of target applications is one of the common factors taken into consideration in works researching mechanisms for efficient decision making on which cloud service provider(s) to use \citep{hosseinzadeh2020service}.
This also appears to be the case for the related problem of optimizing the selection of services from potentially across service providers, commonly known as cloud service composition \citep{amato2016multiobjective,vakili2017comprehensive}.
Managing the cost (and energy consumption) is also identified as one of the focus points of architecting cloud-based software as discussed in the survey of \cite{chauhan2017architecting} on the topic.

\section{Study Design}
\label{sec:design}

In this section, we elaborate on the methods employed to achieve the objective presented in Section~\ref{sec:intro}.
In particular, we discuss the derived research questions, the required data and its collection, and how the data is analyzed to provide the necessary answers.

\subsection{Research Questions}

To explore developers' cost awareness and how it manifests in project repositories, we define three main questions:

\begin{description}
    \item[\rqn{1}] What kind of relevant information can we extract from commits on IaC artifacts?
    \item[\rqn{2}] How can we augment this information further based on issues raised in the respective repositories?
    \item[\rqn{3}] How can we organize this information so we can gain deeper insights from it?
\end{description}

In absence of any prior study establishing a link between activities in repositories and cost awareness, we define these questions on a purely exploratory base.
We strive to find information related to actions taken on cloud-based software projects on the basis of impacting their deployment cost.
In an initial exploration, we seek stronger evidence of such actions and, thus, focus on changes to code connected with acknowledged impact to cost by means of commit messages (\rqn{1}). 
Next, we expand the search scope for discussions that may not incur in changes but that are relevant nevertheless (\rqn{2}).
In the case of this study, we explore entries in the issue trackers of projects that were identified in the previous research question.
At this phase, it is also relevant to understand how the topics of discussion differ (if at all) compared to the information on commit messages.
This information can guide future research and development efforts (see Section~\ref{sec:implications}).

The information collected in the previous RQs can inform future research and practitioner decisions.
However, this `rawer' format of the data may often require examining the dataset in more depth to make connections between the identified core concepts and more informed decisions.
Thus, it is instrumental to understand how the knowledge evolving from the information extracted in the previous research questions can be structured for this purpose (\rqn{3}).

\subsection{Case Selection}
\label{sec:design-case}

As with previous studies on mining software repositories, we direct our efforts to open source repositories.
The reasoning here is that using such an open collection of repositories minimizes the selection bias on our side and therefore increases the robustness of our possible findings.
Among other qualities, we seek a source that can provide a \emph{sizable} and \emph{diverse} population.
Also, from a population that meets the quality criteria, we must find the \emph{breadth} and \emph{depth} of the data that one can derive.
As a source of repositories, we choose GitHub mainly due to the volume and diversity of software projects that are available in it.
Moreover, we aim at maximizing the data pool while maintaining a systematic and repeatable approach and, therefore, GitHub's search features and API are instrumental. 

We also already put forward in the introductory section our particular interest in investigating projects that use cloud orchestrators.
Since we seek to identify evidence of developers' discussions over cloud infrastructure matters (in this case, cost), it is natural to narrow down our search scope to projects that use IaC.
The commit messages and issues involving descriptor files have the potential to bring up concerns we are interested in.
For the purposes of this study, we choose to work with Terraform\footnote{\url{https://www.terraform.io/}} as it is one of the most notable and widely adopted orchestrators \citep{deCarvalho2020}.
Terraform is known for providing an open-source version, cloud services compatibility, interface accessibility and mature API \citep{deCarvalho2020}.

We clarify that there are other viable options, Cloudify\footnote{\url{https://cloudify.co/}} being a well known one, marginally behind Terraform in terms of performance \citep{Kovcs2017}.
Hyperscalers are also offering their own IaC solutions, with AWS' CloudFormation being a particularly popular one.
In principle, therefore, we could execute this study with artifacts of multiple cloud orchestrators taken into account.
However, the complexity of analyzing multiple platforms was deemed prohibitive since this kind of study is a resource-demanding endeavor as-is.
Moreover, we found the number of projects on GitHub mentioning Cloudify (876 repositories with around 79K commits) to be significantly fewer than that of projects mentioning Terraform (171K with around 1M commits).
Cloudify has some of its features locked behind a paywall \citep{Kovcs2017}, potentially turning away many small time and open-source developers, which might explain this difference in numbers.
Furthermore, provider-specific cloud orchestrators would need a deeper understanding on our part of the cloud services being used which would detract from the focus of the study.
Nevertheless, we acknowledge the relevance of investigating other platforms in future work (see Section~\ref{sec:implications}).

In summary, the cases under consideration for this study comprise \textit{GitHub projects that use Terraform as their cloud orchestrator, and address matters related to cost in commit messages and issues discussions}. In the following we describe how we actually apply the latter criterion.

\subsection{Data Collection and Analysis}
\label{sec:design-datacollection}

Following best practices in evidence-based software engineering \citep{wohlin2012ebse}, we characterize the population of this study in terms of \emph{unit of analysis}.
The units of this study are commits or issues in repositories that contain evidence of cloud cost awareness.
For each valid unit, we extract the triplet \verb|<unit-id;unit-content;label>|, where:
(a) \texttt{unit-id} refers to the repository and commit hash or issue id,
(b) \texttt{unit-content} is the commit message or issue text, and
(c) \texttt{label} is a descriptor highlighting the main theme(s) derived from \texttt{unit-content}.

\begin{figure*}
\centering
\includegraphics[width=.89\textwidth]{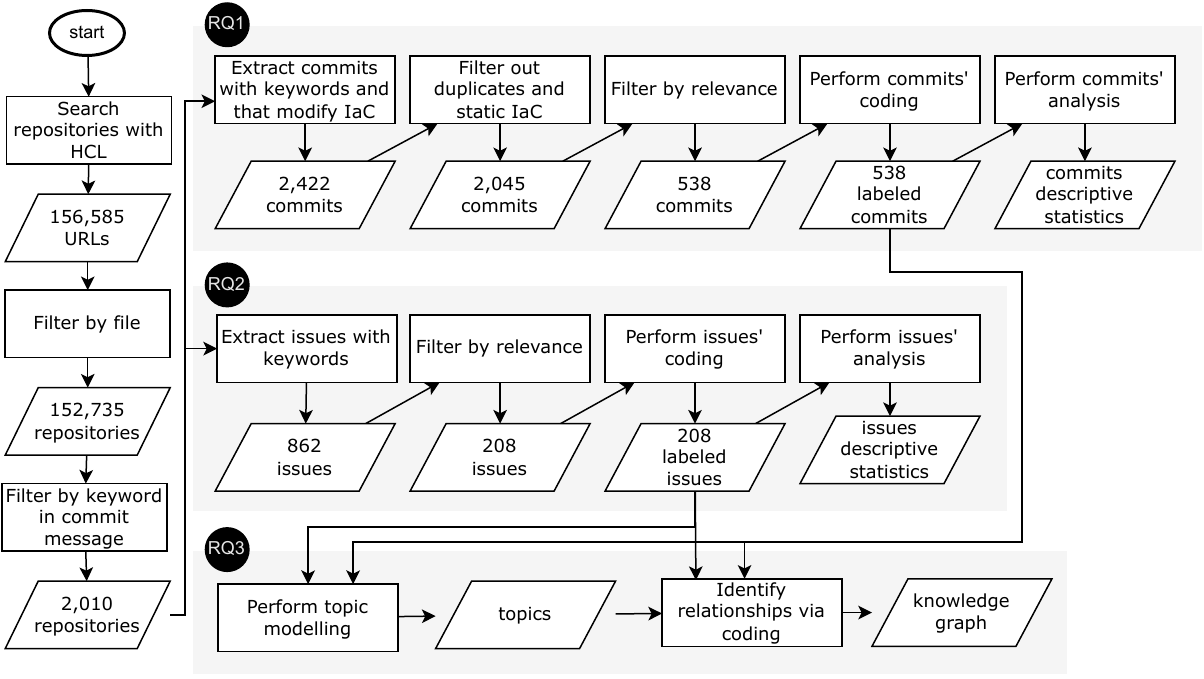}
\caption{Steps of data collection and analysis.}
\label{fig:study_design}
\end{figure*}

The process to extract and analyze the units is summarized in \autoref{fig:study_design}.
Locating Terraform descriptor files is relatively straightforward since they are by default written in the HashiCorp Configuration Language (HCL), a language that is indexed by GitHub, and carry the \texttt{.tf} or \texttt{.tf.json} extension as per the Terraform documentation \citep{HCL}.
Furthermore, Terraform's first release was in 2014, allowing us to constrain our search further for repositories after this date.
To automate the interaction with GitHub's API, we use PyGitHub\footnote{\url{https://pygithub.readthedocs.io/}}. 
Therefore we set PyGitHub to search, on a day-to-day basis\footnote{This was done to avoid the limitations in the amount of results by the GitHub API per request.}, until the end of May 2022 (when the data collection for this study took place), for repositories that contain HCL files created after 2014.
The search returns a candidate set of $156\,585$ repository links.
Removing repositories from this set that do not include \texttt{.tf} or \texttt{.tf.json} files reduces this set to $152\,735$ repositories for further consideration. 

We then use a list of keyword stems to match against commit messages and further search in this set for cost-awareness.
However, it is worth noticing that since there are no previous works on the same topic we cannot reuse their keywords list and we have to come up with our own.
After some piloting, and considering the research questions we aim to answer, we decide on the following keyword stems\footnote{Treating them as stems means that \texttt{expens} for example will match against \texttt{expense}, \texttt{expenses}, \texttt{inexpensive}, and \texttt{inexpensively}.} (in alphabetical order):
\begingroup
\setlength{\leftmargini}{0.15in}
\begin{quote}
    \texttt{bill}, \texttt{cheap}, \texttt{cost}, \texttt{efficient}, \texttt{expens}, and \texttt{pay}. 
\end{quote}
\endgroup

Using this list, we instruct PyDriller \citep{spadini2018pydriller} to process all commit messages in the candidate repositories for the presence of one or more these keywords and we output the repository name, commit hash and message.
PyDriller is a Python library used for analyzing Git repositories.
We chose this tool for its easy-to-use API and broad adoption by the MSR community, which we interpret as an additional sign of quality.  

After this step, a much more manageable set of $2\,010$ repositories containing $6\,116$ potentially related commits is identified as a result.
This set of commits and their respective repositories serve as input to collect the data and perform the analysis for each research question.

\subsubsection{\rqn{1}}
\label{sec:design-datacollection-rq1}

To answer the first research question, we must identify what kind of information related to cloud cost management can be extracted from the commits.
Out of this set of $6\,116$ commits, $377$ are from forked repositories already in the set.
After filtering out these commits and those that do not modify any Terraform files, we are left with $1\,162$ repositories and $2\,045$ related commits.
The selected commits are then inspected manually to decide their actual relevance.
For that, each commit message is checked by two researchers and validated by a third one.
Any conflicts are resolved by the entire team in consolidation meetings.
The output of this process results in the identification of $538$ relevant commits.
The selected units come from $434$ distinct repositories.

\begin{figure*}
\centering
\includegraphics[width=\textwidth]{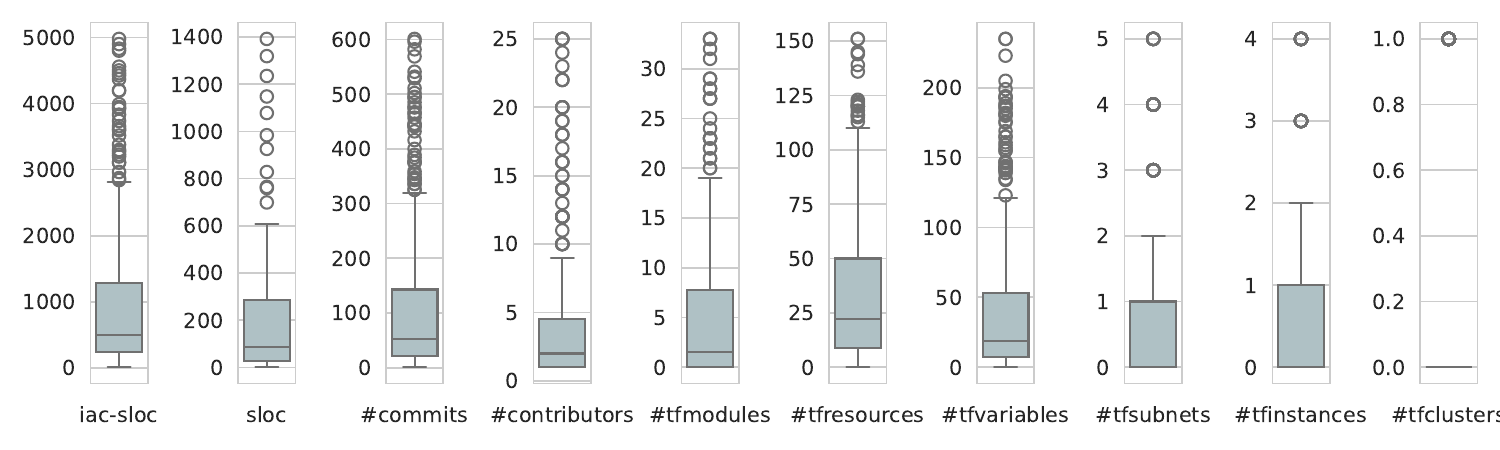}
\caption{\rqn{1} repository demographics (90\textsuperscript{th} percentile).}
\label{fig:repo-demographics-commits}
\end{figure*}

In \autoref{fig:repo-demographics-commits}, we characterize the repositories in terms of IaC artifacts' size (\texttt{iac-sloc}, max. $89$K), source code size (\texttt{sloc}, max. $1\,210$K), number of commits (\texttt{\#commits}, max. $17$K), contributors (\texttt{\#contributors}, max. $596$), number of Terraform modules (\texttt{\#tfmodules}, max. $1.4$K), resources (\texttt{\#tfresources}, max. $2.5$K), variables (\texttt{\#tfvariables}, max. $5$K), subnets-related definitions (\texttt{\#tfsubnets}, max. $32$), instances-related definitions (\texttt{\#tfinstances}, max. $36$), and number of Terraform clusters-related definitions (\texttt{\#tfclusters}, max. $21$).
The max values are not visible in \autoref{fig:repo-demographics-commits} since we needed to trim the top $10\%$ values of each variable to better visualize the plot.
We also note that $65\%$ ($284$) of the population comprise repositories that contain (and manage) IaC artifacts only (i.e., they do not contain source code).
These characteristics suggest that the population is varied with a tendency for repositories maintained by few contributors and concentrated in repositories of more Terraform and other IaC files than source code.
This is possibly an indication of developers taking over a ``Terraform expert'' role in teams that become responsible also with dealing with the operational cost aspect of each project.

We then proceed to collect what kind of information related to cloud cost management can be extracted from these commits.
This is primarily a manual task, to be performed using open coding, a form of inductive coding \citep{corbin2014qualitative}.
For this task, each data point (for this RQ, commit message) is labeled first by two researchers and validated by a third.
The labels refer to central ideas in the discussion and their characteristics.
Any conflicts (incl. disagreements) are resolved by all five researchers in a consolidation meeting.
In this meeting, we verify the rationale for the codes and analyzed the evidence again (i.e., the complete commit message), and argued to a consensus.
We did not have cases where a consensus was not met after this process.

As it will be discussed further in Section~\ref{sec:results}, all collected units are analyzed to identify relevant characteristics such as the prevalence of the various labels and distribution of the units among repositories.
In addition, meta-information such as the \texttt{unit-id} is used to analyze the repositories, e.g., regarding the number of source lines of code (SLOC) and number of contributors.

\subsubsection{\rqn{2}}
\label{sec:design-datacollection-rq2}

For the second research question, we wish to explore what kind of additional information related to cloud cost awareness can be extracted from entries in issue trackers that adds to that extracted from commit messages.
As shown in \autoref{fig:study_design}, we start from the list of $1\,339$ repositories obtained from the first filtering of commits based on keyword.
We use this list because discussion may take place before actual changes happen (as demonstrated by modification to Terraform files), and we would like to capture them too.
Also, this set of repositories allows us to narrow down the population to an amount of issues that we can feasibly extract from GitHub since they already contain some evidence that cost may be a concern for the project.

To extract the issues, we provide GrimoreLab's Perceval \citep{duenas2018perceval} with the repository owner username,
the repository name, and a GitHub API token.
Perceval is a Python-based tool that can collect data from various sources, including issue trackers.
We chose Perceval for its versatility and validation within the MSR community.
The tool then returns us a list of issue objects that contain every single detail pertaining the issue.
We then extract the issue objects that contain one or more of the defined keywords in the title, body or any of the comments; this results in an initial set of $862$ entries.

\begin{figure*}
\centering
\includegraphics[width=\textwidth]{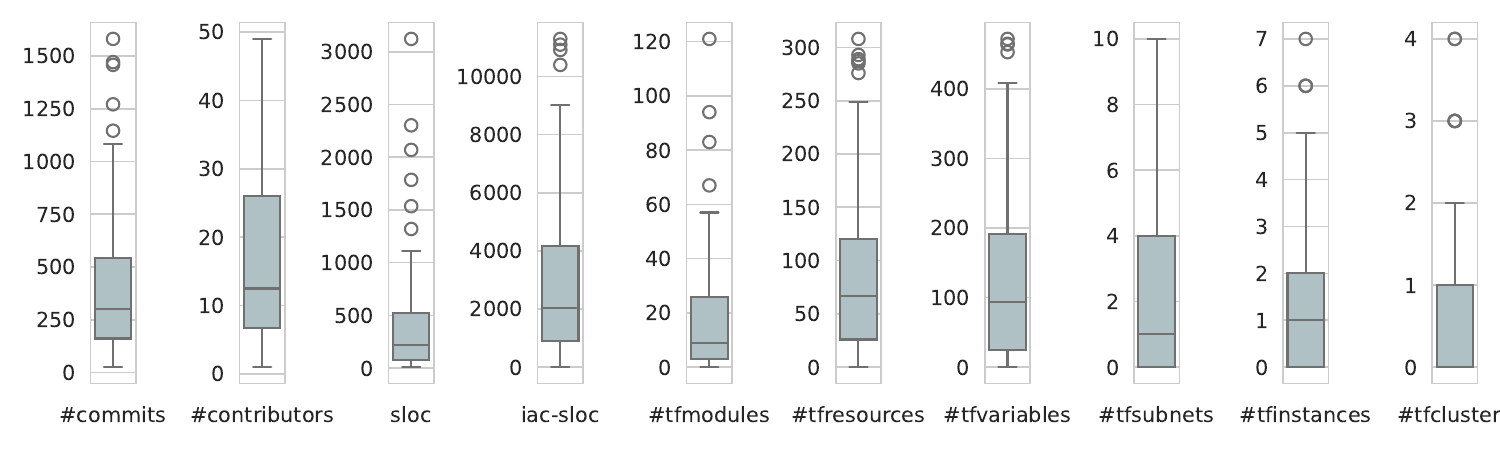}
\caption{\rqn{2} repository demographics (90\textsuperscript{th} percentile).}
\label{fig:repo-demographics-issues}
\end{figure*}

Next, we apply a process similar to that described for \rqn{1} to filter for relevant issues and then label them.
Since issues may contain long discussions, the coding is focused on the context, i.e., the sentence where the keyword appears and the surrounding ones (when needed).
If multiple keywords appear on the same issue and they refer to different contexts, multiple checks are performed.
We note that the labels are applied to whole issues based on evidence found in any of the content elements (i.e., title, body or comments).
This process culminates in the selection of $208$ units belonging to $89$ distinct repositories.
In \autoref{fig:repo-demographics-issues}, we characterize the repositories in terms of the same metrics as in \autoref{fig:repo-demographics-commits}: \texttt{iac-sloc} (max. $43$K), \texttt{sloc} (max. $1\,486$K), \texttt{\#commits} (max. $6$K), \texttt{\#contributors}, (max. $156$), \texttt{\#tfmodules} (max. $373$K), \texttt{\#tfresources} (max. $646$K), \texttt{\#tfvariables} (max. $3.5$K), \texttt{\#tfsubnets}, (max. $39$), \texttt{\#tfinstances} (max. $34$), and \texttt{\#tfclusters}, max. $27$).
We again trimmed the top $10\%$ values of each variable to better visualize the plot.
This population of repositories is considerably smaller than that collected for \rqn{1}, and contains a lower amount of projects that contain (and manage) Terraform files (only $44\%$).
Nevertheless, the descriptive statistics summarized by \autoref{fig:repo-demographics-issues} suggest a fair distribution of data points, with higher averages (by approx. a factor of two).

Similar to \rqn{1}, the collected units are analyzed to identify relevant characteristics such as the prevalence of the various labels and distribution of the units among repositories.
Moreover, the results are compared to those obtained from \rqn{1}, to reflect on the value added by exploring issues.
Both of these issues are to be discussed further in Section~\ref{sec:results}.

\subsubsection{\rqn{3}}
\label{sec:design-datacollection-rq3}

In the final research question, we want to examine the dataset in more depth and make contextual connections between the identified core concepts.
This process is done in two steps, starting with topic modeling~\cite{Blei2010ProbabilisticTM} and followed by a second coding activity, culminating in the creation of a knowledge graph which combines the results of the previous steps.
We describe these steps in the following.

Topic modeling is a statistical learning tool that is well-suited for abstracting connections between words (into topics) from a corpus of documents~\cite{Blei2010ProbabilisticTM}.
We apply Latent Dirichlet Allocation (LDA)~\cite{Blei2010ProbabilisticTM}, a popular topic-modeling technique used in a number of MSR studies e.g.~\cite{zimmerle2022mining,alamin2021empirical,chen2012explaining,hindle2011automated}.

Before applying LDA, a number of steps must be undertaken for cleaning and preparing the corpus.
GitHub content contains a mixture of Markdown and HTML code scattered throughout the text that we mined.
Thus, we first convert any markdown syntax to HTML, remove the content in \textit{pre}, \textit{code} and \textit{blockquote} elements, and remove the markup and URLs.
We note that we remove code snippets because we are interested in developers' discussions and the code may bias the LDA to find irrelevant topics.
Next, we prepare each document using Stanza's \citep{qi2020stanza} neural network NLP pipeline\footnote{\url{https://stanfordnlp.github.io/stanza/pipeline.html}}.
In particular, we tokenize it into sentences containing lists of tokens, extract part-of-speech (PoS) tags of each token, and lemmatize them.
The latter step is especially relevant to improve the validity of the bag of words used for topic modeling.
From the prepared tokens, we filter the PoS tags that may contain relevant words, i.e., nouns, adjectives, adverbs and verbs.
We also remove tokens according to a stop-word list that we built by merging the Gensim's \citep{rehurek_lrec} list for English 
with terms deemed irrelevant for our analysis.
We perform the same steps for both commits' and issues' text.

With the prepared corpus, we use Gensim to convert it into a bag of words, build a TF-IDF (term frequency–inverse document frequency) model and use the two to create LDA models.
To build an LDA model, we must find a suitable number of topics ($K$), as it impacts the granularity of the results \citep{abdellatif2020challenges,han2020what}.
The quality of the model is also affected by other hyperparameters, such as $\alpha$ (referring to document-topic density) and $\beta$\footnote{In Gensim, the parameter `eta' refers to $\beta$.} (referring to topic-word density) are some of the most relevant \citep{campbell2015latent,treude2019predicting}.
Following the related literature \citep{reboucas2016empirical,alamin2021empirical,zimmerle2022mining}, we experimented with the ranges $K=\{5,6,\dots,34,35\}$, $\alpha=\beta=\{50/K,0.01\}$.
We also varied the \emph{chuncksize} (number of documents for each training mini-batch)\footnote{\url{https://radimrehurek.com/gensim/models/ldamodel.html}} $S=\{1,2,4,8,...,1024\}$ as it can yield positive impact on the model~\cite{hoffman2010online}.
Although the quality of the model is ultimately assessed by us, we relied on the coherence \citep{abdellatif2020challenges,alamin2021empirical} and perplexity \citep{treude2019predicting,zimmerle2022mining} metrics to narrow down the number of candidate models to be manually inspected.
Finally, we train models using 100 iterations for hyperparameter exploration, and use 1000 iterations to train the models selected for manual inspection.
After this process, we settled with $K=12$, $\alpha=50/K$, $\beta=0.01$, and $S=32$ for the model based on the commits dataset model; and $K=5$, $\alpha=0.01$, $\beta=50/K$, and $S=2$ for the model based on the issues dataset.

Next, we aim to build a knowledge graph by making informed connections between relevant words.
The set words come from both the coding performed for \rqn{1} and \rqn{2} as well as from the interpretation of the topics of both models.
To interpret the topics, we used the open card sorting technique \citep{abdellatif2020challenges,zimmerle2022mining} and analyzed the top words of a topic via a random sample of documents dominated by it \citep{ahmed2018what}.
The sample size varied from 10 to 20 documents per topic, depending on the number of documents connected to a given topic (we aimed for 5-10\% of the number of documents connected to the topic).
Three researchers applied the technique, with the other two researchers validating the results and resolving disagreements.
Finally, the relationship between words is defined by applying axial coding and selective coding \citep{corbin2014qualitative} on the preexisting labels (from \rqn{1} and \rqn{2}) and words from topics.
Axial coding is a combination of inductive and deductive coding with the goal of relating codes (e.g., finding emerging categories).
Selective coding is a similar process, but to find the core set of codes and categories.
Group categories and relationship between labels can be identified from a topic (if a word can describe the topic well) and from manual inspection of labels and their linked documents.

\section{Results}
\label{sec:results}
The presentation of the results follows the RQs defined in Section~\ref{sec:design}.

\subsection{Cost Awareness in Commits (\rqn{1})}
\label{sec:results-rq1}

\begin{figure}
\centering
\includegraphics[width=.49\textwidth]{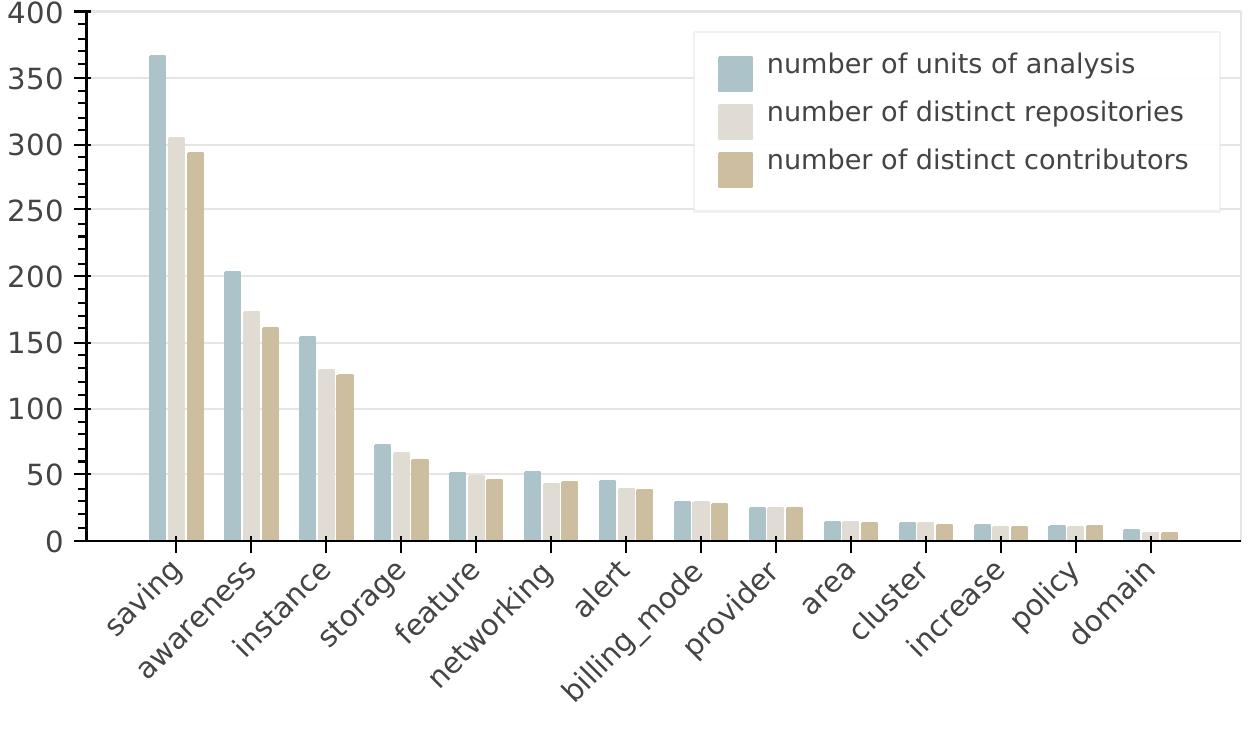}
\caption{Coding demographics of commits.}
\label{fig:coding-demographics-commits}
\end{figure}

As a result of the coding process, we obtain a set of 14 distinct labels related to the $538$ commits.
Each commit has between one and five labels assigned to it based on the message content.
\autoref{fig:coding-demographics-commits} lists the collected labels, and shows their recurrence among units (i.e., commits) and distinct repositories, and number of distinct contributors associated with them.
We provide a description of each label in Table~\ref{tab:codedesc} and elaborate on the most recurrent ones in the following.

\begin{table*}
    \caption{Label descriptions}
    \label{tab:codedesc}
    \centering
    \begin{tabularx}{\linewidth}{>{\ttfamily}lX}
    \hline
    {\normalfont\textbf{label}} & \textbf{description} \\
    \hline
    alert & text expressing concerns related to billing alarms enforcing an upper threshold on costs.\\
    area & text expressing concerns related to server or instance geographical location.\\
    awareness & text simply mentioning concerns with cost (without necessarily implying action).\\
    billing\_mode & text expressing concerns related to the type of billing plan being used (e.g., on-demand for development or normal plan for production).\\
    cluster & text expressing concerns related to cluster configuration.\\
    domain & text expressing concerns related to domain name system and IP addresses.\\
    feature & text expressing concerns related to various features such as logging, load balancers or usage of third party libraries.\\
    increase & text expressing concerns related to increase in cost due to a change.\\
    instance & text expressing concerns related to computational instances (e.g., Amazon AWS t2.micro) used in the deployment.\\
    networking & text expressing concerns related to networking configuration.\\
    policy & text expressing concerns related to the implementation of general rules to prevent excessive charges.\\
    provider & text expressing concerns related to choosing a service providers (e.g., Amazon, Azure, Google).\\
    saving & denotes mentioned changes made to save costs.\\
    storage & text expressing concerns related to storage solutions (e.g., Amazon gp2) used in the deployment.\\
    \hline
    \end{tabularx}
\end{table*}

The most popular label by all metrics is \code{saving}.
A notorious example from the dataset is:
\begin{tcolorbox}
    \texttt{Removed the default use of detailed monitoring. (\#17) * Reduces CloudWatch costs for metrics by 80\%}\\
    \textit{blinkist/terraform-aws-airship-ecs-cluster
    (commit hash: d7aa6599)}
\end{tcolorbox}
\noindent This example showcases the fact that monitoring solutions provided by the cloud service providers might offer deep insights into the billing of their services, but are also incurring expenses for their usage, as with any other cloud service.
This being the most common label indicates not only awareness on behalf of the developers, but also specific cost reducing actions as an effect of this awareness.
As a matter of fact, approximately $70\%$ of all commits as shown in \autoref{fig:coding-demographics-commits} document concrete actions to save cost.
Characterizing the types of actions taken is outside of the scope of this work, but can be easily achieved by further processing the commits in the dataset.

The next most popular label is \code{awareness}, which does also imply action, e.g.:
\begin{tcolorbox}
    \texttt{nat gateway is verry {\normalfont [sic]} expensive}\\
    \textit{stealthHat/k8s-terraform
    (hash: 681a3f8b)}
\end{tcolorbox}
\noindent
This particular example identifies a well known issue with Amazon Web Services' NAT Gateway service with respect to cost accruing easily out of control that is even the subject of online memes and frequently recurring Twitter threads\footnote{See e.g.\ \url{https://twitter.com/quinnypig/status/1440301033314349062}}.

Finally, the label \code{instance} understandably figures among the top recurring labels.
An example from the dataset is:
\begin{tcolorbox}
    \texttt{Move from m4.large to m5.large. The new gen have more CPU and are cheaper}\\
    \textit{alphagov/govuk-aws
    (hash: 6cfda6ad)}
\end{tcolorbox}
\noindent
In this respect, this label can be located anywhere between the previous two ones: it can identify awareness and intention of action at the same time.
On a related note, it is worth pointing out that, although less prevalent, we also identified information explicitly discussing cost increases due to a prior change (in roughly $2\%$ of the units).

Altogether, we notice developers' consciousness of consequences of decisions in deployment.
Moreover, labels such as \code{instance} and \code{storage} point to specific aspects of the deployment that are or can be tuned to manage cost.
From \autoref{fig:coding-demographics-commits}, we cannot infer or speculate over which of such aspects are more often treated this way.
However, considering the configuration options offered by platforms such as Terraform, the data suggest a broad understanding of these options.
At the same time, the number of repositories in our dataset compared to the total amount of projects using Terraform may also suggest that only a small percentage of developers are interested or aware of the cost-saving possibilities in IaC configuration.
We note that this observation is also speculative as it requires further investigation to e.g.~discard test or template projects.
Further insights can be gained by analyzing the dataset in more depth for e.g.~characterizing the types of repositories with respect to the labels used in their commits, identifying the relation between labels and contributors and so on.
This kind of analysis is left as future work and as part of a call to the wider community, as discussed below in Section~\ref{sec:implications}.

\subsection{Cost Awareness in Issues (\rqn{2})}
\label{sec:results-rq2}

\begin{figure}
\centering
\includegraphics[width=.49\textwidth]{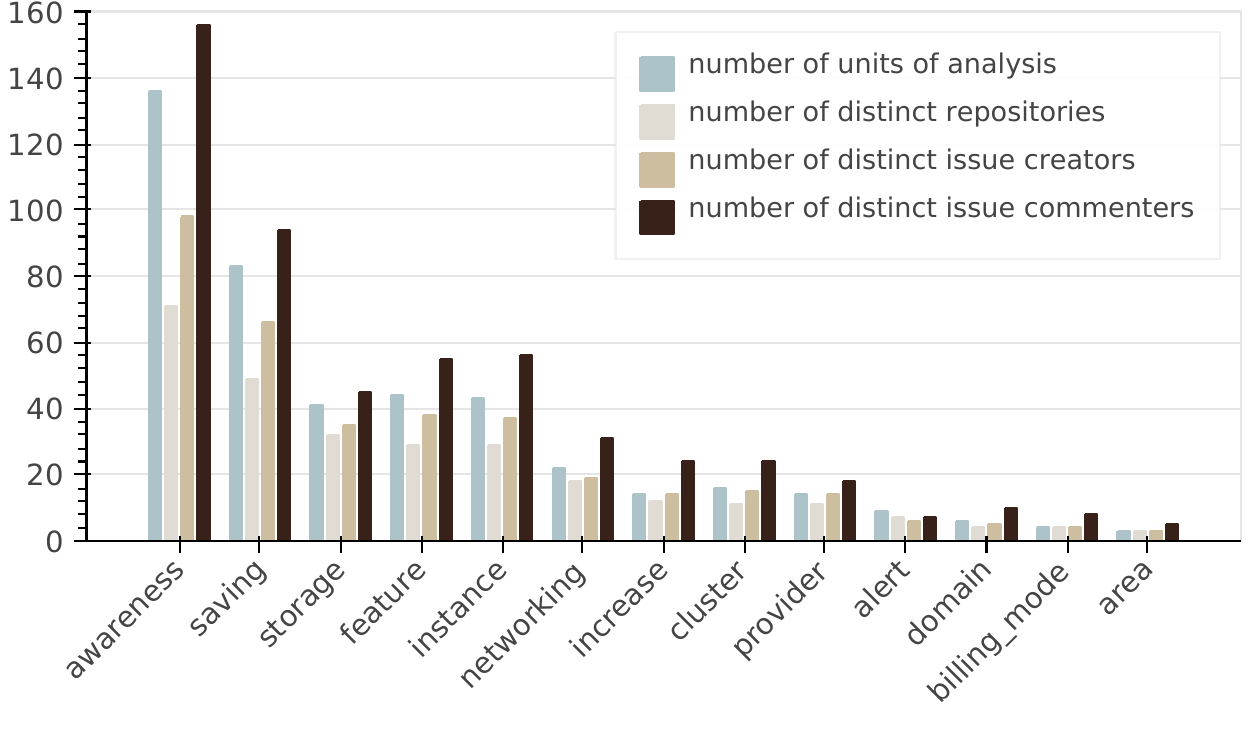}
\caption{Coding demographics of issues.}
\label{fig:coding-demographics-issues}
\end{figure}

The coding of the $208$ issues resulted in recognizing labels we identified in the commit-units of analysis, without adding new ones.
We note that one label, namely \code{policy}, was not identified among issues.
Each issue discussion has between one and four labels assigned to it based on the text in the title, body and comments.
\autoref{fig:coding-demographics-issues} describe the labels in terms of their recurrence among units (i.e., issues) and distinct repositories, and number of distinct creators and commenters.

At a first glance, we notice considerable similarity between \autoref{fig:coding-demographics-commits} and \autoref{fig:coding-demographics-issues}.
We explored this observation further by inspecting the content of a random sample of units from both datasets that are tagged with the same label.
For that, we selected 50 commits and 20 issues, i.e. aiming for a representativeness of 10\%.
In general, we notice that issues contain more information around the cost-related matter at hand, which is expected since they essentially provide a discussion forum.
More importantly, we found the extra amount of information to be often related to decision-making around the cost matter.
Actors may present hints on the current configuration of the deployment, the alternatives for change, the rationale and even potential feature requests.
For example, see the following two samples from different issues showing the depth of information that can be potentially extracted:
\begin{tcolorbox}
    \texttt{By having a personal deployment, we're free to experiment and research without any limitation. The drawback is that it implies a cost for the cloud provider.Alternatively, we could imagine a `sandbox.qhub.dev` or `alpha.qhub.dev` or whatever}\\
    \textit{Quansight/qhub (issues: \#924)}
    \hfill\textbf{label}: \texttt{provider}
\end{tcolorbox}

\begin{tcolorbox}
    \texttt{But the LBs (HTTP(s) and TCP) don't work because they only have the default/main worker pool as target pool, and in my setup its size is 0. So I am kinda paying for Global FW rules that have no use and I can't delete them because they will get created again in the next `terraform apply`.}\\
    \textit{poseidon/typhoon (issue: \#558)}
    \hfill\textbf{label}: \texttt{networking}
\end{tcolorbox}

Thus, while commit messages are commonly more compact and may report cost-changing actions, issues may shed light on the decision-making process that developers undergo before applying a cost-affecting change.
An example of this contrast can be seen in the following units both labeled as \code{instance}:

\begin{tcolorbox}
    \texttt{Change code to use the cheaper r4.xlarge instances type.}\\
    \textit{cisagov/cyhy\_amis (hash: 4e67a501)}
\end{tcolorbox}

\begin{tcolorbox}
    \texttt{It would be really great if the new `t4g` instance, which are even cheaper than `t3.nano`, would be supported as well}\\
    \textit{Guimove/terraform-aws-bastion (issue: \#124)}
\end{tcolorbox}

\noindent While the commit message communicates the change for a cheaper instance, the issue hints on the current configuration and expresses the wish for a new feature (i.e., support of a different instance).
We note that this example regards different repositories. 
We tried to find connections between commits and issues but our dataset does not contain any direct links to issue in commit messages.
An alternative to explore this avenue is to study pull requests, which is outside of our scope but mentioned in our research agenda (see Section~\ref{sec:implications}).

\begin{figure}
\centering
\includegraphics[width=.49\textwidth]{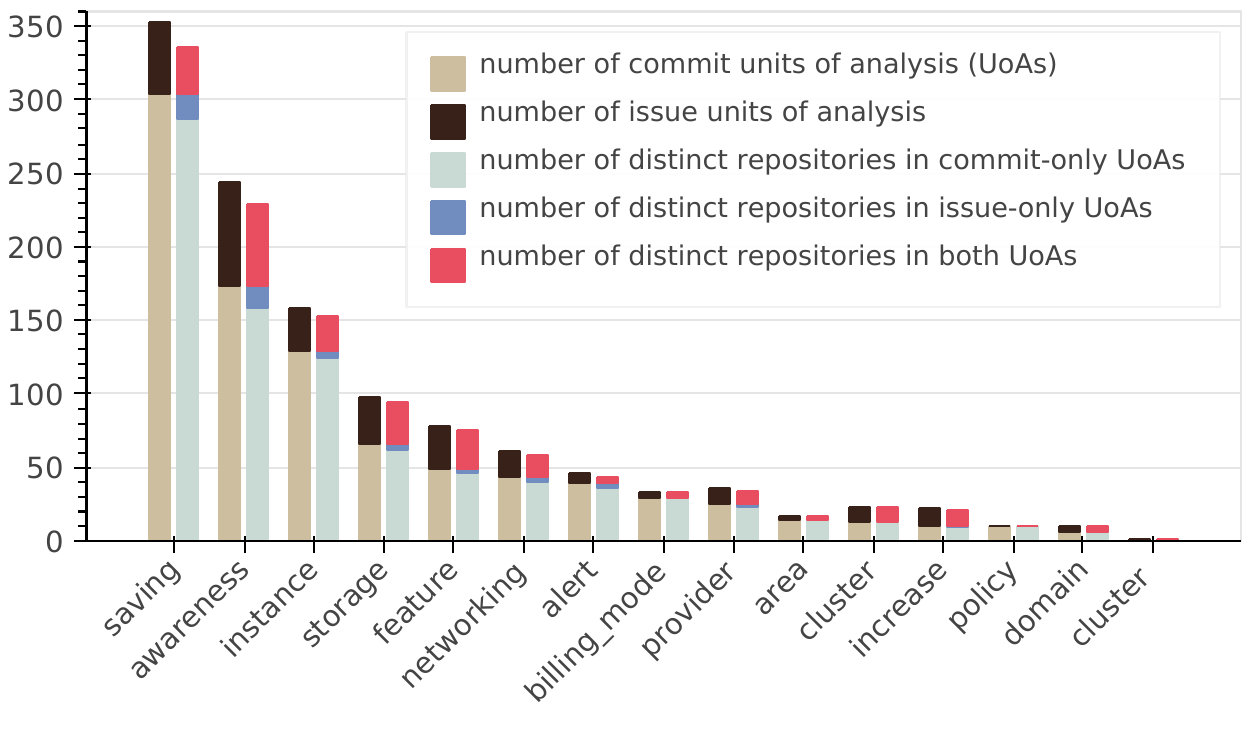}
\caption{Coding demographics of the combined dataset.}
\label{fig:coding-demographics-all}
\end{figure}

Continuing with the analysis, we notice some changes in the order of the labels in terms of recurrence.
\autoref{fig:coding-demographics-all} shows a comparison of the two sets of units and helps to visualize their differences.
\code{awareness} is more recurrent than \code{saving} among issues, which might be related to our observation that issues can be a more prominent platform for decision-making.
On a similar note, units labeled with \code{increase} are also more recurrent among issues (compared to commit units).
This might also be in line with the nature of issues, in this case, reporting or acknowledging cost increase.
The example below depicts a situation of a seemingly unintentional increase.

\begin{tcolorbox}
    \texttt{Pods for some core services have migrated over to high-memory nodes, which have a much higher cost that the general nodes. I tried killing the pod hoping it would restart on a different node, but usually it just restarts on the same node.}\\
    \textit{Quansight/qhub (issue: \#321)}
\end{tcolorbox}

From \autoref{fig:coding-demographics-all}, we also observe that, despite the lower number of repositories among issue-units of analysis, the majority ($54$ out of $89$, and an average of $85\%$ per label) are unique to them.
In conclusion, despite the similarities of the assigned labels, the results suggest that the information extracted from issues can complement that from commits.
In particular, issues can provide more knowledge about the decision-making surrounding cost management in the deployment of cloud systems configured with IaC.

\subsection{Knowledge Organization (\rqn{3})}
\label{sec:results-rq3}

From the initial coding performed for \rqn{1} and \rqn{2}, we obtained a set of 14 labels.
Moreover, as we delved in the units' content to further understand what kind of information can be extracted, we found evidence of connections and additional relevant themes.
For example, in the unit regarding \textit{alphagov/govuk-aws} (hash: 6cfda6ad) presented in Section~\ref{sec:results-rq1}, we notice a clear link between \code{instance} and \code{saving}, as well as indication of what motivates the change (i.e., more CPU).
This particular unit had already been tagged with both labels, but our labels are not fine-grained to the point of providing  lower levels of detail.

We note that one main reason for not using finer-grained labels during coding was to avoid explosion of labels that, although meaningful, might not have been ultimately relevant (i.e., not recurrent enough) and could risk the quality of the procedure.
Our goal is to guide future research and practitioners' decisions by providing a more generalizable and actionable knowledge based on developers' experience.
Thus, we now take a step back to collect more relevant labels and establish connections between them.
As described in Section~\ref{sec:design-datacollection-rq3}, we started by modeling topics from our dataset, and then proceed to perform axial and selective coding based on the topics and the labels identified in our dataset.

During the investigation of \rqn{2}, we established that the text in both commit messages and issues' content is varied but complementary in nature.
So we aimed to create one model per set of units to avoid the risk of not identifying relevant topics.
As a result of our topic modeling, we identified 12 topics from commit units and five topics from issue units.
The used hyperparameter configurations (see Section~\ref{sec:design-datacollection-rq3}) yielded the most promising results, but not all topics were useful for our purposes.

\begin{figure*}
    \centering
    \includegraphics[width=\linewidth]{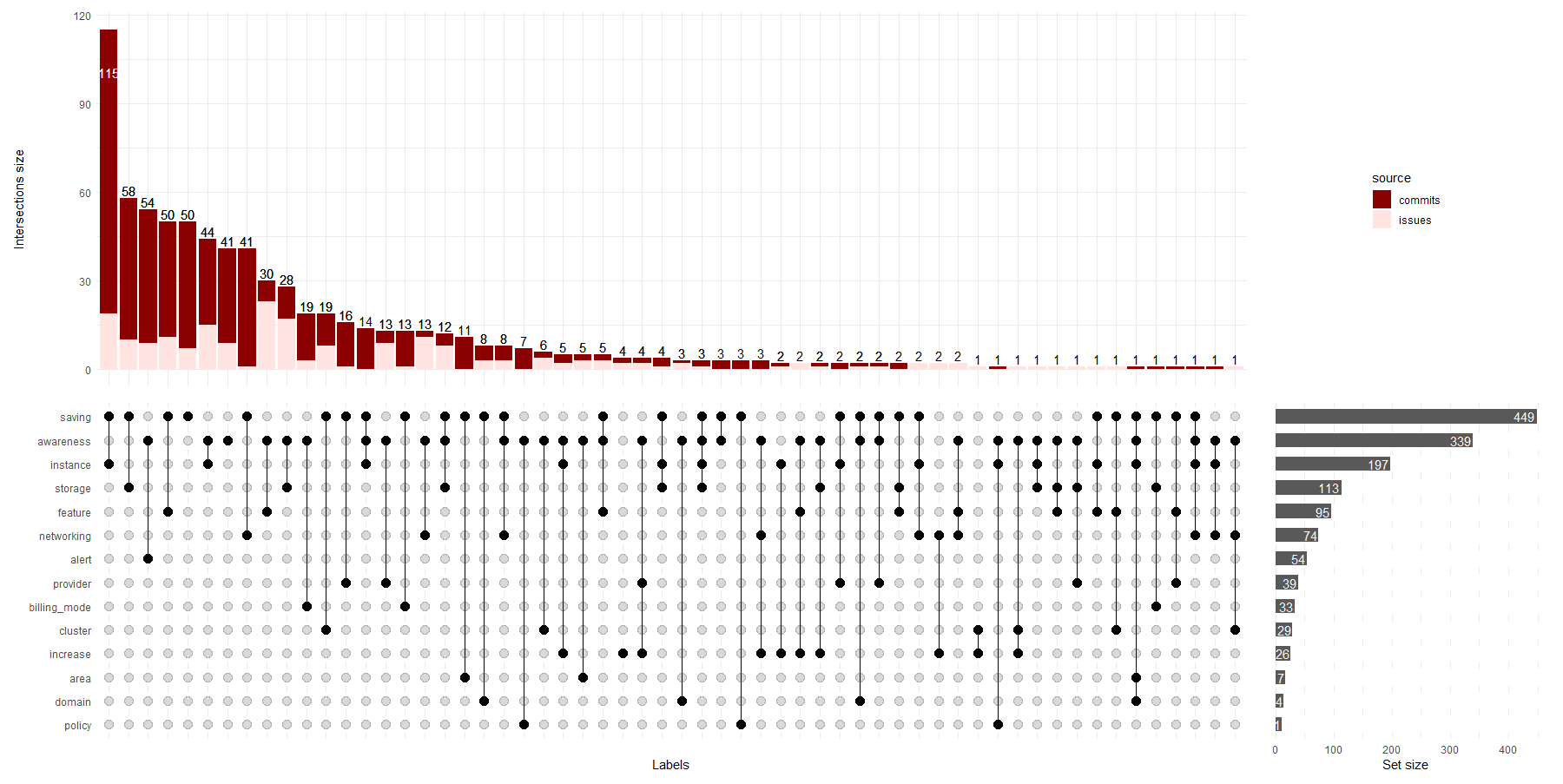}
    \caption{UpSet plot showing the occurrences and co-occurrences of topic labels in commit messages and issues discussions as label intersections resulting from the open coding in \rqn{1} and \rqn{2}.}
    \label{fig:coding-intersections-commits}
\end{figure*}

In particular, only one topic derived from issues was considered.
The relevant topics mention \textbf{cost-related terms} (e.g., `cheap', `expensive', `budget', `waste') and \textbf{actions} (e.g., `change', `move', `add', `test', `upgrade') associated with various \textbf{properties of the deployment}, both general (e.g., `VM', `storage', `disk', `machine') and specific (e.g., `dynamodb', `CPU', `NAT', `EC2', `RAM').
Some of the connections revealed through the topics are already observable in our coding for \rqn{1} and \rqn{2}, in the form of co-occurring labels as summarized by
\autoref{fig:coding-intersections-commits}.
The figure presents these co-occurence relationships as an UpSet plot~\citep{lex2014upset}.
The labels (with their frequencies) are shown as rows on the bottom part, and 
the frequency of the various combinations of labels are represented as columns on the top part and described through the connected dots on the bottom part.

\begin{figure}
    \centering
    \includegraphics[width=.98\columnwidth]{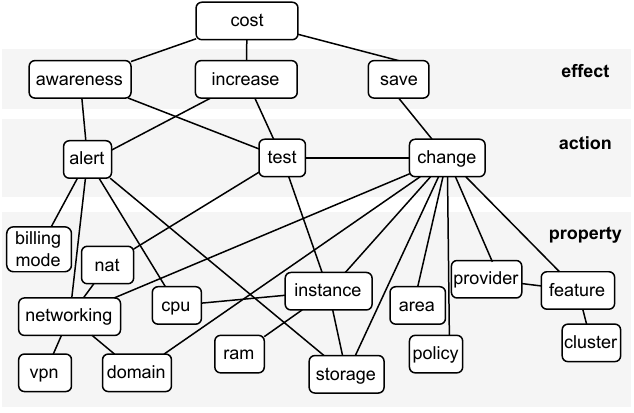}
    \caption{Knowledge graph resulting from the axial and selecting coding; label \texttt{save} replaces the label \texttt{saving} used in open coding.}
    \label{fig:knowledge-graph}
    \vspace{-10pt}
\end{figure}

We then applied axial and selective coding to aggregate terms (from topics and labels) and inspect the relationships in the units.
At the end of this process, we created the knowledge graph depicted in \autoref{fig:knowledge-graph}.
The graph compiles three main levels of information: effects on cost, actions related to an effect, and the properties of the deployment that are considered for the action.
The edges reflect the most significant connections we found and subsequently confirmed in the dataset.

The most recurrent relationships identified during the open coding (for \rqn{1} and \rqn{2}) are also represented in the graph.
Moreover, new links between preexisting labels were found based on topics (e.g., between \code{increase} and \code{alert}) and new, more specific, terms appeared (e.g., `CPU').
We highlight that, although seeing specific terms in one context is what led us to add the associated general label (e.g., \code{instance} for `CPU'), the fact that the specific term appears in a topic is a strong indicative that it is prominent.
Such particular cases prompted us to include the specific term to graph and connect it to the related terms and the more general one.

After identifying their relationships, some of the topics identified for the previous question can be interpreted as specific pain points that developers are clearly concerned about: choices concerning networking, instance and provider selection, and the applicable billing mode (\textit{property} level in \autoref{fig:knowledge-graph}). Other topics signify specific actions taken to address these concerns, or awareness that these actions could/should be taken to avoid unnecessary costs: setting or removing alerts, testing for one or more of these concerns (e.g.\ the use of VPN), or changing something in the Terraform file towards directly dealing with these concerns (\textit{action} level in \autoref{fig:knowledge-graph}). All these actions, both affected and intended to be affected can be characterized based on their desired outcome: increasing awareness, dealing with increasing costs, or produce savings (\textit{effect} level in \autoref{fig:knowledge-graph}).

In conclusion, the information aggregated in the knowledge graph serves as a summary of relevant cost-related concerns and actions when deploying cloud-based applications using Terraform.
It also introduces the subjects one may expect to find in our dataset in terms of both content and context.
Altogether, it opens the door to extract deeper insights and inform research and deployment decisions, as discussed below.

\section{Results Triangulation}
\label{sec:triangulation}

To verify and strengthen our findings, we sought to investigate the extent to which the identified related topics are also discussed among developers outside GitHub. In particular, in this section, we analyze the discussions of developers in Stack Overflow\footnote{\url{https://stackoverflow.com/}} (SO) and aim at the triangulating the observations from this empirical work with those presented in Section~\ref{sec:results-rq3}. We start by presenting the data collection procedure, followed by the performed analysis, and concluding with the results and comparison against our previous observations.

\subsection{Data Collection}

The data collection entailed the extraction of relevant discussions from SO. For that, we used key datasets from the September 12, 2023 data dump, namely, posts\footnote{\url{https://archive.org/download/stack-exchange-data-dump-2023-09-12/stackoverflow.com-Posts.7z}}, comments\footnote{\url{https://archive.org/download/stack-exchange-data-dump-2023-09-12/stackoverflow.com-Comments.7z}} and change histories\footnote{\url{https://archive.org/download/stack-exchange-data-dump-2023-09-12/stackoverflow.com-PostHistory.7z}}, totaling 57.3 GiB. As the goal of this process is to triangulate the results of the study on GitHub reponsitories, we specifically targeted discussions about Terraform, rooted in questions containing tags with the string `terraform' in the post's metadata. Thus, we started by filtering questions fitting the mentioned criteria, which led to a dataset of $19\,139$ questions stored as JSON files.

We gathered the comments and post histories associated with each question and updated the respective JSON file to include them. We then extracted the answers linked to each question, identified through their parent post IDs, and added the comments and post histories similarly to the questions. The answers (and associated data) were then added to the question files, creating the final version of the dataset of $19\,139$ questions and their associated answers, complete with their respective comments and post histories.

\subsection{Data Analysis}

In the analysis, we aimed at investigating whether or not the concepts depicted in \autoref{fig:knowledge-graph} are also present and prominent in SO discussions. For that, we started with filtering cost-related questions by searching the body, title, comments and history of all questions and associated answer(s) for the same keyword stems used for filtering commit messages (see Section~\ref{sec:design-datacollection}): \texttt{bill}, \texttt{cheap}, \texttt{cost}, \texttt{efficient}, \texttt{expens}, and \texttt{pay}.

Next, we searched the same fields mentioned above of each filtered question for the properties documented in our knowledge graph (see \autoref{fig:knowledge-graph}). In this process, we augmented each question JSON file with a list of filtered sentences that contain one or more of these properties. Finally, two of the authors manually inspected a representative sample of the questions where properties were found with the goal of establishing the accuracy of this process. 
We note that the generated dataset and used scripts are publicly available \citep{supplementary_material_so}.

\subsection{Results and Discussion}

The initial filtering (using keywords stems) returned a total of 491 questions, i.e., approx. 2.6\% of all questions on SO with `terraform' in any of the tags. Although the sample size may seem small compared to the entire population of Terraform-related questions, we note that this is double the sample size compared to that of GitHub repositories containing cost-related commits (1.3\%, $2\,010$ out of $152\,735$ containing Terraform artifacts; see Section~\ref{sec:design-datacollection}). This proportional increase may be further indicative of developers' involvement with and attention to cost-related decisions.

\begin{figure}
\centering
\includegraphics[width=.49\textwidth]{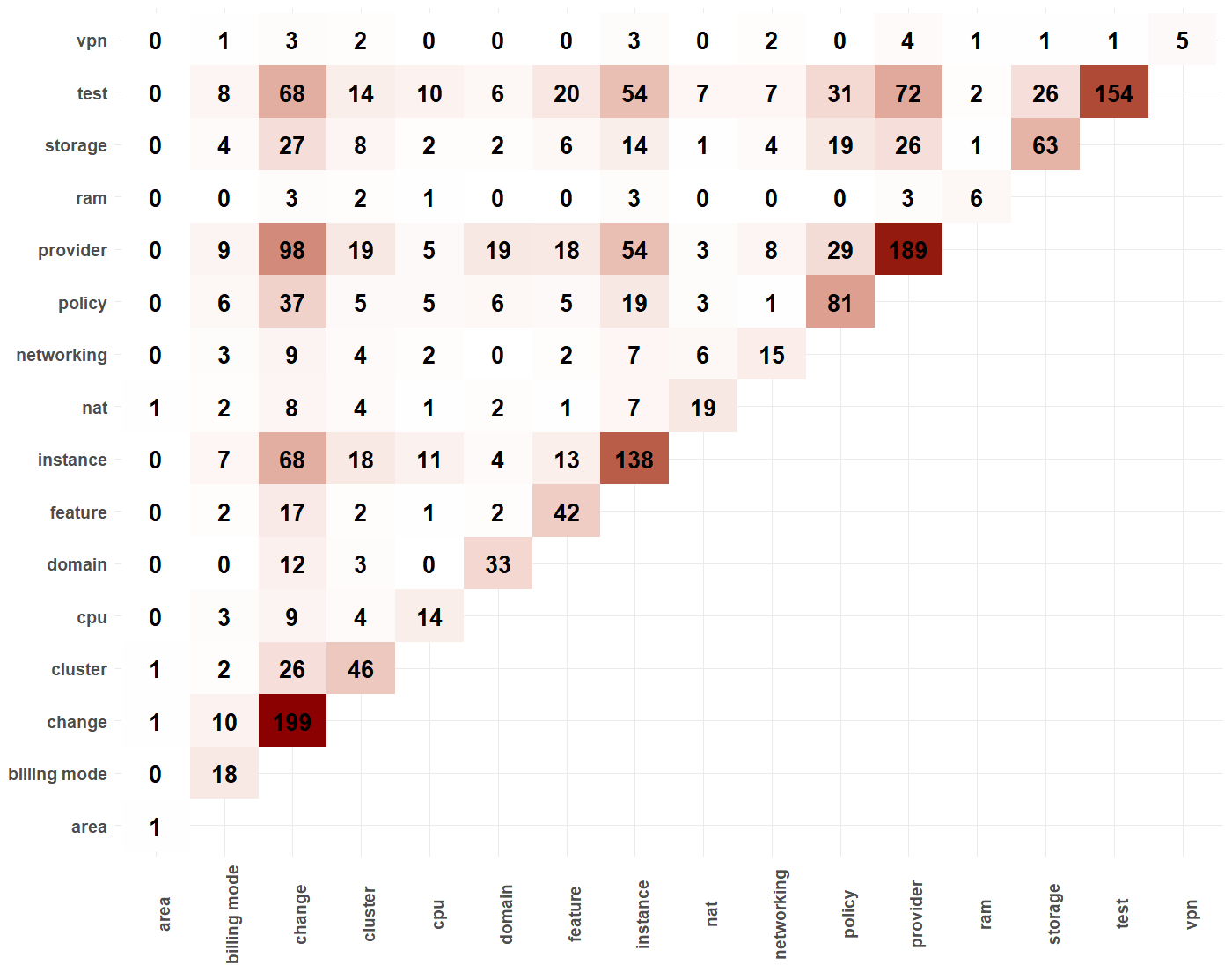}
\caption{Co-occurrences of action and property concepts from the knowledge graph of \autoref{fig:knowledge-graph} in Stack Overflow questions and their associated answers.}
\label{fig:cooccurrences}
\end{figure}

Upon inspecting these questions and their answers for actions and properties from our knowledge graph, we learned that 451 (approx. 92\%) of them mention one or more of these concepts. In the diagonal of \autoref{fig:cooccurrences}, we show the exact number of questions and associated answers where each concept was present as a term. While properties such as \texttt{area} appear only once, other properties such as \texttt{provider} and \texttt{instance} (189 and 138 occurrences, respectively) but also actions such as \texttt{change} (199) and \texttt{test} (154) are very popular. The same question and its associated answers might contain a reference to more than one such concept. From these questions' JSON file, we then extracted $3\,934$ sentences containing one or more of these concepts. 

Extracted sentences showcase developers sharing their experiences in dealing with insidious cost-inducing technicalities in using existing services or asking for advice in this direction:

\begin{figure*}
    \centering
    \includegraphics[width=.9\textwidth]{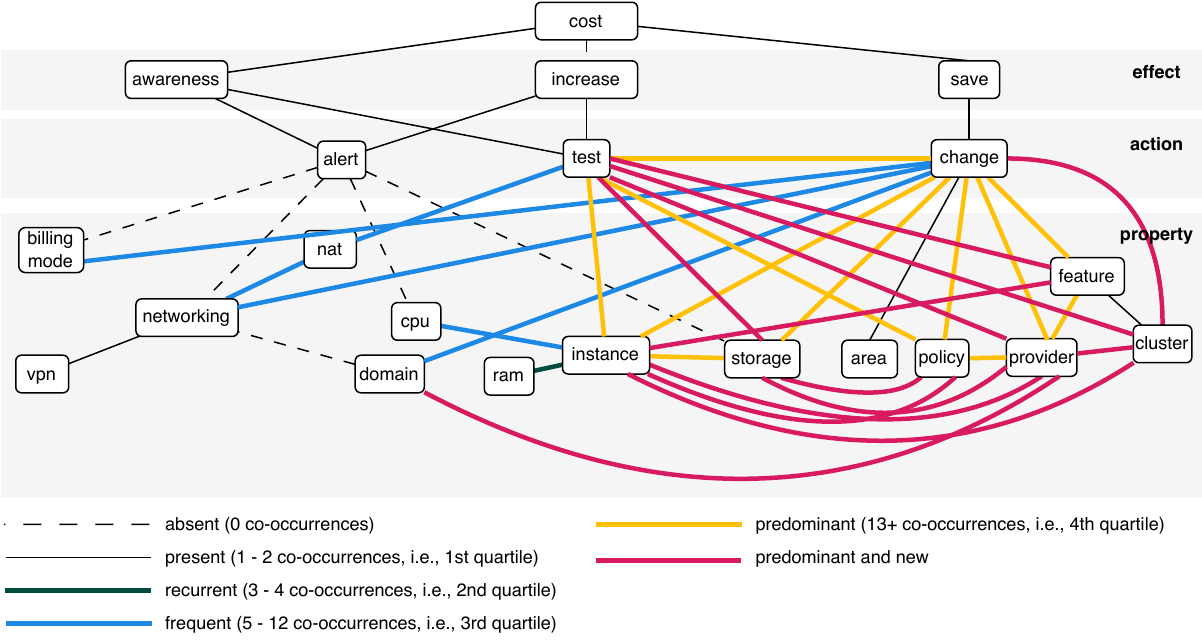}
    \caption{Knowledge graph updated based on the co-occurrences of concepts in the Stack Overflow posts.}
    \label{fig:knowledge-graph-update}
    \vspace{-10pt}
\end{figure*}

\begin{tcolorbox}
    \texttt{Azure by itself discusses the challenge you are addressing and recommends using ""data lifecycle"" for reducing the storage costs. [...] After you enable blob versioning for a storage account, every write operation to a blob in that account results in the creation of a new version.}\\
    \textit{post id: \#72881767}
    \hfill\textbf{label}: \texttt{storage}
\end{tcolorbox}

\begin{tcolorbox}
    \texttt{I am trying to reduce the cost of my AWS infrastructure deployed using Terraform for a Django app. I have 2 public subnets and 2 private subnets and in the subnets I deploy NAT gateways and elastic ips. all works but is expensive. }\\
    \textit{post id: \#74422443}
    \hfill\textbf{label}: \texttt{nat}
\end{tcolorbox}

Developers even identify cost-related features requested to be added to Terraform as an answer to existing problems:

\begin{tcolorbox}
    \texttt{Also savings plans have largely replaced reserved instances for now so I'd be tracking this issue \url{https://github.com/hashicorp/terraform-provider-aws/issues/10785} if you are interesting in support for using Terraform to manage savings plans.}\\
    \textit{post id: \#48751593}
    \hfill\textbf{label}: \texttt{provider}
\end{tcolorbox}

Adding sentences where the initial keywords (i.e.\ cost, bill, etc.) are mentioned provides additional insights:
\begin{tcolorbox}
    \texttt{Since licensing of SQL Server is expensive, I want to switch it off at least for the night. }\\
    \textit{post id: \#76474230}
    \hfill\textbf{label}: \texttt{expense}
\end{tcolorbox}
\begin{tcolorbox}
    \texttt{I have multiple databases running in each environment which are charging me a lot of cost each month, so i wanted to downscale the DTUs to some lower count during non-working hours, again during working hours DTUs to be upscale back to actual DTUs count, it should happen automatically as per time settings every single day.}\\
    \textit{post id: \#74538381}
    \hfill\textbf{label}: \texttt{cost}
\end{tcolorbox}
Some of these discussions point towards the need for architectural refactoring as a way of addressing cost-related concerns:
\begin{tcolorbox}
    \texttt{If you need zero cost during idle time, should you go with serverless with new design?}\\
    \textit{post id: \#56530721}
    \hfill\textbf{label}: \texttt{cost}
\end{tcolorbox}

Aided by these highlighted sentences, we proceeded to manually inspect a representative sample of 37 questions\footnote{\url{https://www.calculator.net/sample-size-calculator.html?type=1&cl=95&ci=5&pp=2.6&ps=491&x=Calculate}}, considering that we selected 2.6\% of a population of $19\,139$ questions with Terraform tags. 
During the inspection, we aimed to verify whether: (1) the question indeed contained a cost-related discussion (either as the main problem or as the element of a description or argument), and (2) one of the properties is an essential element of the identified cost-related text. As a result, we found that all inspected questions were true positives. These combined observations show that the knowledge graph cannot only be used to identify actual pain points faced by cloud application developers, but it also points to the most recurrent ones. 

Based on the co-occurrences identified in \autoref{fig:cooccurrences}, we proceeded to investigate if and how the knowledge graph could be augmented. We note that the frequency of the observed co-occurrences varies greatly while also being positively skewed, i.e., many low co-occurrences (min.:~1, \textit{Q}1:~2, \textit{Q}2:~5, \textit{Q}3:~12, max.:~98). Following our goal of representing the most meaningful information in the knowledge graph, we processed the co-occurrence frequencies as follows:
\begin{enumerate}
    \item mark preexisting graph edges that were not observed as \texttt{absent};
    \item classify the observed co-occurrences into \texttt{present} ($f\leq 2$), \texttt{recurrent} ($2<f\leq 4$), \texttt{frequent} ($4<f\leq 12$) and \texttt{predominant} ($12<f$) according to the four quartiles;
    \item mark preexisting graph edges observed in the data according to their frequency class (established in the previous step); and
    \item add new graph edges for predominant co-occurrences that were not present in the previous version.
\end{enumerate}

The resulting knowledge graph is depicted in \autoref{fig:knowledge-graph-update}. We note that the markings explained in the figure legend are aimed at helping the reader visualize the process. The augmented graph comprise all pre-existing edges with the addition of the new edges for predominant co-occurrences. Looking at the differences and co-occurrences frequencies, while many relationships between concepts are also observed in this dataset, we learn two main lessons:
(a) the cost-related topics revolve mainly around the ability to try out different configurations (i.e., \texttt{test}) and updating configuration (i.e., \texttt{change}), also meaning that creating system alerts is not frequently discussed; and
(b) discussions are more concrete in the sense that they get more specific about deployment matters such as features and benefits of providers, e.g., appropriate instances and policies.
The latter point, in turn, led to a number of new relations being identified between ``low level'' concepts in the graph, e.g., between \texttt{instance} and \texttt{policy} or \texttt{cluster}.

\section{Implications \& Future Work}
\label{sec:implications}

The presented findings have implications for both practitioners and academic researchers.
With respect to the former, it becomes clear that cost awareness should actually be present, if it is not already, throughout the development of cloud-based applications.
Our dataset contains multiple examples of developers rushing to adapt their deployment configurations to deal with prohibitive costs, or intentionally designing their deployment with the clear intention of avoiding them, particularly when specific cloud services are involved.
More importantly, there are specific pain points and actions that can be taken for reducing costs in these cases and lessons learned by other developers to be extracted by investigating our curated dataset.
Looking at the questions posted on Stack Overflow on the topic, and the corresponding answers, it becomes clear that having such knowledge in advance at their disposal would help practitioners avoid common mistakes and pitfalls in managing their operational expenses in the cloud.

With respect to the researcher community, our findings demonstrate that there is indeed empirical evidence of software developers being aware of the cost of their choices with respect to deploying their software in the cloud.
This evidence is corroborated by the number of posts on Stack Overflow discussing the topics we identified as important for cost awareness.
While this conclusion is the product of processing only a specific type of artifacts in open source repositories and user forums, there is no reason to make us believe that there is no further evidence to be uncovered when other types of artifacts, or even the software source code in the identified repositories in the dataset is examined.
The findings of this study are therefore a call for further studies on cost awareness during software development.

More specifically, the following research items can be pursued by starting with our existing dataset:
\begin{itemize}
    \item[\ding{229}] Provide a finer-grain analysis of the collected evidence looking at e.g.\ the commit contributors and type of projects involved. Combined, for example, with a practitioner's survey or interviews, it can shed light on how different development teams and organizations deal with cost-related issues.
    \item[\ding{229}] Collect and correlate cost awareness evidence from the pull requests (PRs) of the identified repositories. PRs can provide the missing link between commits and issues and offer further insights into the cost management practices.
    \item[\ding{229}] Extend the evidence search to other cloud orchestrator solutions, both provider-agnostic (e.g.~TOSCA\footnote{\url{https://docs.oasis-open.org/tosca/TOSCA/v1.0/TOSCA-v1.0.html}}) and -specific ones (AWS CloudFormation\footnote{\url{https://aws.amazon.com/cloudformation/}}), and compare the findings. Especially for the latter, GitHub might not necessarily be the best data source for this purpose, with the repositories of (large) organizations and enterprises with mature cloud presence over the years being much more attractive sources of data.
    \item[\ding{229}] Apply natural language processing and other machine learning techniques such as sentiment analysis to gain further insights into the reasoning of the developers. Sentiment analysis in particular has been shown to be a mixed bag when software is concerned~\cite{lin2022opinion}, but with the training of the analyzer focused on cloud orchestrator artifacts and their associated commit messages and issue discussions, it might be possible to get better results through the tighter scope. 
    \item[\ding{229}] Look for similar evidence in other types of artifacts, for example other configuration files, and/or in closed source repositories, e.g.\ of large software-intensive organizations and enterprises. This search can go beyond cloud-based software, or at least public cloud deployments, and incorporate also the wealth of DevOps tools available for automating software deployment and management. 
    \item[\ding{229}] Look also in the source code of repositories, starting with the ones already containing cost-aware orchestration artifacts as easier to reach targets. This is an obvious extension of the current work, but will need a wide experience with multiple programming languages and platforms on behalf of the researchers.
    \item[\ding{229}] Use the extracted posts from Stack Overflow to extend the list of topics related to cost awareness, and update the knowledge graph of \autoref{fig:knowledge-graph} accordingly.
    \item[\ding{229}] Organize the collected information into reusable knowledge concerning the best practices of managing the operational expenses of cloud-based software. In a sense, this would be the outcome of this research agenda with the most impact to the wider community.
\end{itemize}

We strongly believe that this is only the first study of many to come on this particular topic.

\section{Threats to Validity}
\label{sec:ttv}
Like any other empirical study in software engineering, this work's validity is also threatened (and these threats mitigated) in several ways.
The main threats to this work are discussed in the following.
\newline

\noindent
\textbf{External Validity}: Regarding external validity, the population in our data sets (from GitHub and Stack Overflow) may not represent all possible cost-related discussions in cloud projects.
The inclusion of repositories hosted in other (closed-source) platforms, or that use other cloud orchestrators, or that use no cloud orchestrators, could lead to the identification of new discussions.
The inclusion of discussion forums other than Stack Overflow could also lead to new discussions.
However, our decisions were carefully considered with the aim of ensuring data quality, diversity and quantity while keeping the execution feasible for the available human resources.
Furthermore, our goal was not to collect all possible evidence, to begin with, but rather to see if there is any evidence available.
That said, the results triangulation (between GitHub- and Stack Overflow-based discussions) helps mitigate threats to external validity.
\newline

\noindent
\textbf{Construct Validity}: The selection criteria, and most noticeably the defined keywords may threat the construct validity of our work.
We mitigated this threat by piloting and testing our selection criteria, and considering the knowledge of both academic and industrial domain experts.
Furthermore, the coding activity is naturally open to subjectivity and inconsistencies.
To mitigate this threat, we followed a well-established process, 
and introduced an extra final step to consolidate the knowledge and systematically discuss the labels.
Also, the topics derived from applying LDA may not fully represent their content.
We mitigate this threat by manually inspecting the units of analysis during the investigation of \rqn{3}, which was part of the axial coding and selective coding.
This threat is also mitigated by the triangulation presented in Section~\ref{sec:triangulation}.
That said, we acknowledge that the Stack Overflow study itself suffers from its own threats to construct validity, mainly related to selection of appropriate data points, i.e., pertaining to cost-related discussion of Terraform-based deployments.
The mitigation strategies in this case entail the use of Terraform-related tags that have been assigned by users, on top of using previously-validated keywords for searching cost-related entries within the filtered posts.
\newline

\noindent
\textbf{Reliability}: Finally, to mitigate threats to the reliability of the study, we have described the data acquisition process in as much detail as possible.
More importantly, the dataset curated through our efforts is publicly available \citep{supplementary_material} together with the scripts to aid the replication the data collection and topic modeling tasks. We also share the dataset and scripts used for the results triangulation with Stack Overflow questions in a separate package \citep{supplementary_material_so}.

\section{Conclusions}
\label{sec:conclusions}

Managing the operational expenses of deploying software in the cloud is a major challenge for organizations. However, how practitioners approach this topic has so far been treated by the literature in an anecdotal and therefore non systematic manner. Consequently, and as a first step, in this work we investigated whether there is (empirical) evidence of software developers being aware of the operational expenses of deploying and delivering software in the cloud, and if yes, then what kind of information can be extracted from this evidence and how this information can be organized for further study. 

Given the wide scope and previously unexamined nature of these questions, at least from an MSR perspective, we started tackling them by focusing on cloud orchestrator descriptor files.
We chose repository mining as our methodology, and designed and executed the first such study searching for evidence in open source code repositories on GitHub that contain Terraform orchestration artifacts, a very popular provider-agnostic IaC tool.

Our search was shown to be successful, insofar as it actually allowed us to retrieve and organize in a dataset not only evidence of cost awareness by software developers, but also of specific actions being taken as a result.
More specifically, with respect to extracting information from commit messages in the selected repositories (\rqn{1}), our findings show that the most popular topics dominating the developer discourse is not limited to being aware of the potential or actual cost of deploying and operating the system in the cloud. 
Developers seem also to not only be taking concrete actions to minimize this cost, but also to avoid excessive charging to occur when e.g.\ using specific cloud services and/or offerings within them.
Processing cost-related issues from the same repositories (\rqn{2}) did not reveal any additional information in terms of identified topics of discussion. 
It did however offer further insights into the decision-making process entailed in managing cloud costs that can be pursued further in future work.

Enriching and organizing the extracted information from the previous steps into a knowledge graph (\rqn{3}) helped us identify both higher-level, recurring concepts such as awareness, and specific pain points such CPU and RAM (sizes) in the deployment and operation of cloud-based software that dominate the developers' discussions.
The follow-up triangulation with Stack Overflow Terraform discussions that address cost concerns corroborates these pain points and their prevalence in the spectrum of discussion topics.

Finally, based on that evidence and the limitations of this work we developed a list of future research items which both provides us with a clear roadmap for future work, and offers to the wider community an opportunity to develop a new research topic at the intersection of mining software repositories and cloud computing.

\section*{Acknowledgements}
The authors would like to thank Diomidis Spinellis for providing the inspiration for this work.

\printcredits

\bibliographystyle{cas-model2-names}

\bibliography{bibliography}

\end{document}